\documentstyle[11pt,newpasp,twoside,epsf]{article}
\def\edcomment#1{\iffalse\marginpar{\raggedright\sl#1\/}\else\relax\fi}
\reversemarginpar

\begin{document}
\title{The Inner Density Cusp of Cold Dark Matter Halos }
\author{Julio F. Navarro}
\affil{Department of Physics and Astronomy, University of Victoria, 3800
Finnerty Road, Victoria, BC, Canada V8P 1A1}

\begin{abstract}
I report recent results of numerical simulations designed to study the inner
structure of Cold Dark Matter (CDM) halos. This work confirms the proposal of
Navarro, Frenk \& White (NFW) that the shape of $\Lambda$CDM halo mass profiles
differs strongly from a power law and is approximately independent of mass. The
logarithmic slope of the spherically-averaged density profile, as measured by
$\beta=-d\ln \rho/d\ln r$, decreases monotonically towards the center, becomes
shallower than isothermal ($\beta < 2$) inside a characteristic radius,
$r_{-2}$, and shows no evidence for convergence to a well-defined asymptotic
value ($\beta_0$) at the center. The dark mass contained within the innermost
radius resolved by the simulations places strong constraints on $\beta_0$; cusps
as steep as $r^{-1.5}$ are clearly ruled out in our highest resolution
simulations. A profile where the radial dependence of the slope is a simple
power law, $\beta(r)\propto r^{\alpha}$, approximates the structure of halos
better than the NFW profile and so may minimize errors when extrapolating our
results inward to radii not yet reliably probed by numerical simulations. We
compare the spherically-averaged circular velocity ($V_c$) profiles to the
rotation curves of low surface brightness (LSB) galaxies in the samples of de
Blok et al and Swaters et al. The $V_c$ profiles of simulated CDM halos are
generally consistent with the rotation curves of LSB galaxies, but there are
also some clearly discrepant cases. This disagreement has been interpreted as
excluding the presence of cusps, but it may also just reflect the difference
between circular velocity and rotation speed likely to arise in gaseous disks
embedded within realistic triaxial halos.
\end{abstract}

\section{Introduction}
The pioneering efforts of Frenk et al (1988), Dubinski and Carlberg (1991), and
Crone et al (1994), among others, led to the identification of a number of key
features of the structure of dark matter halos assembled through hierarchical
clustering. One important result of this early work concerns the remarkable
similarity (``universality'') in the structure of dark matter halos of widely
different mass. This was first proposed by Navarro, Frenk
\& White (1996, 1997; hereafter NFW), who suggested a simple fitting formula to
describe the spherically-averaged density profiles of dark matter halos,
\begin{equation}
\rho(r)={\rho_s \over (r/r_s) (1+(r/r_s))^{2}}.
\label{eq:nfw}
\end{equation}
A second result concerns the absence of a well defined central ``core'' of
constant density in virialized CDM halos: the dark matter density grows
apparently without bounds toward the center of the halo. 

Subsequent work has generally confirmed these trends, but has also highlighted
potentially important deviations from the NFW fitting formula. In particular,
Fukushige \& Makino (2001), as well as Moore and collaborators (Moore et al
1998, 1999), have reported that NFW fits to their simulated halos (of much
higher mass and spatial resolution than the original NFW work) underestimate the
dark matter density in the inner regions ($r<r_s$). These authors interpreted
the disagreement as indicative of inner density ``cusps'' steeper than the NFW
profile and advocated a simple modification to the NFW formula (referred to
hereafter as the M99 profile) with asymptotic central slope $\beta_0=1.5$.

The actual value of $\beta_0$ is still being hotly debated in the literature
(Jing \& Suto 2000, Klypin et al 2001, Taylor \& Navarro 2001, Power et al 2003,
Fukushige, Kawai \& Makino 2003, Hayashi et al 2003, Navarro et al 2003), but
there is general consensus that CDM halos are indeed ``cuspy''. This has been
recognized as an important result, since the rotation curves of many disk
galaxies, and in particular of low surface brightness (LSB) systems, appear to
indicate the presence of an extended region of constant dark matter density: a
dark matter ``core'' (Flores \& Primack 1994, Moore 1994).

Unfortunately, rotation curve constraints are strongest just where numerical
simulations are least reliable.  Resolving CDM halos down to the kpc scales
probed by the innermost points of rotation curves poses a significant
computational challenge that has been met in very few of the simulations
published to date.  Indeed, rotation curves are generally compared with
extrapolations of the simulation data that rely heavily on the applicability and
accuracy of fitting formulae such as the NFW profile to regions that may be
compromised by numerical artifact.

The theoretical debate on the asymptotic central slope of the dark matter
density profile, $\beta_0$, has also led at times to unwarranted emphasis on the
very inner region of the rotation curve datasets, rather than on an proper
appraisal of the data over its full radial extent. De Block et al (2001), for
example, attempt to derive constraints on $\beta_0$ from the innermost few
points of their rotation curves, and conclude that $\beta_0\sim 0$ for most
galaxies in their sample. However, their analysis focuses on the regions most
severely affected by non-circular motions, seeing, misalignments and slit
offsets, and other effects that limit the accuracy of circular velocity
estimates. It is perhaps not surprising, then, that other studies (van den Bosch
et al 2000, Swaters et al 2003) have disputed the conclusiveness of these
findings. The disagreement is compounded by the results of the latest
cosmological N-body simulations (Power et al 2003, Hayashi et al 2003, Navarro
et al 2003), which find scant evidence for a well defined value of $\beta_0$ in
simulated CDM halos. Given these difficulties, focusing the theoretical or
observational analysis on $\beta_0$ seems unwise. We shall try here to improve
upon previous work by comparing directly the results of our simulations with the
full radial extent of the rotation curves of LSB galaxies.


I report here results from a major computational effort pursued by an
international collaboration bringing together computational cosmology groups in
Victoria (Canada), Seattle (USA), Durham (UK), and Garching (Germany). We have
now completed a series of about 20 simulations of $\Lambda$CDM halos spanning
roughly five decades in mass, from dwarf galaxies with circular velocities $V_c
\sim 50$ km s$^{-1}$ to galaxy cluster systems with $V_c\sim 1500$ km
s$^{-1}$. This series follows a thorough convergence study where the role of all
relevant numerical parameters has been assessed in detail (Power et al 2003). As
a result, we are able to identify in each simulated halo a minimum convergence
radius ($r_{\rm conv}$) beyond which the mass profile is reliably reproduced:
circular velocity estimates are accurate to better than $10\%$ for $r>r_{\rm
conv}$. Each of the simulated halos has several million particles within the
virial radius, allowing for the mass profiles to be traced reliably down to
within $1\%$ of the virial radius.

\section{CDM Halo Density Profiles: Universality and Cusp Structure}

\begin{figure}
\begin{center}
\plottwo{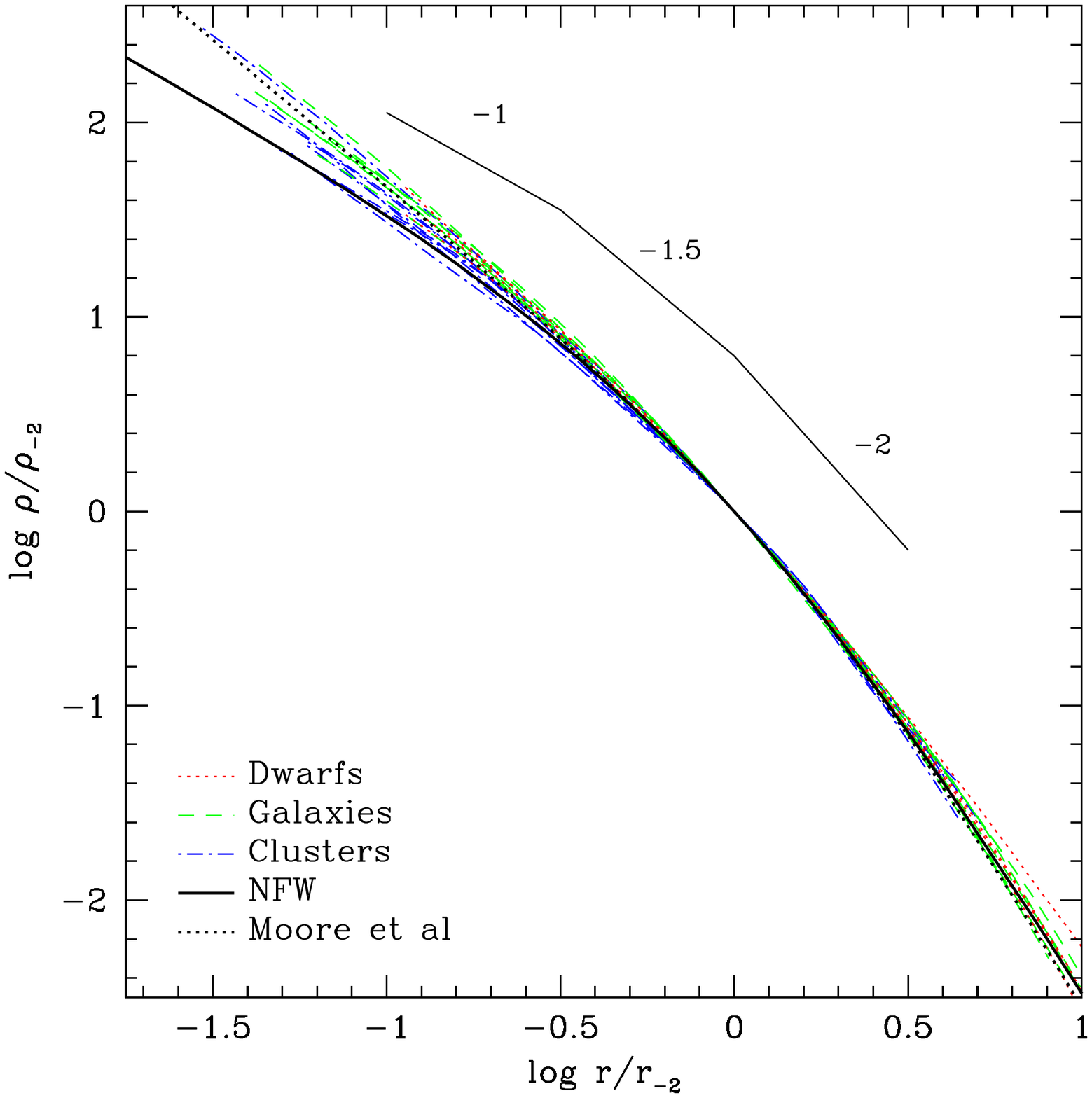}{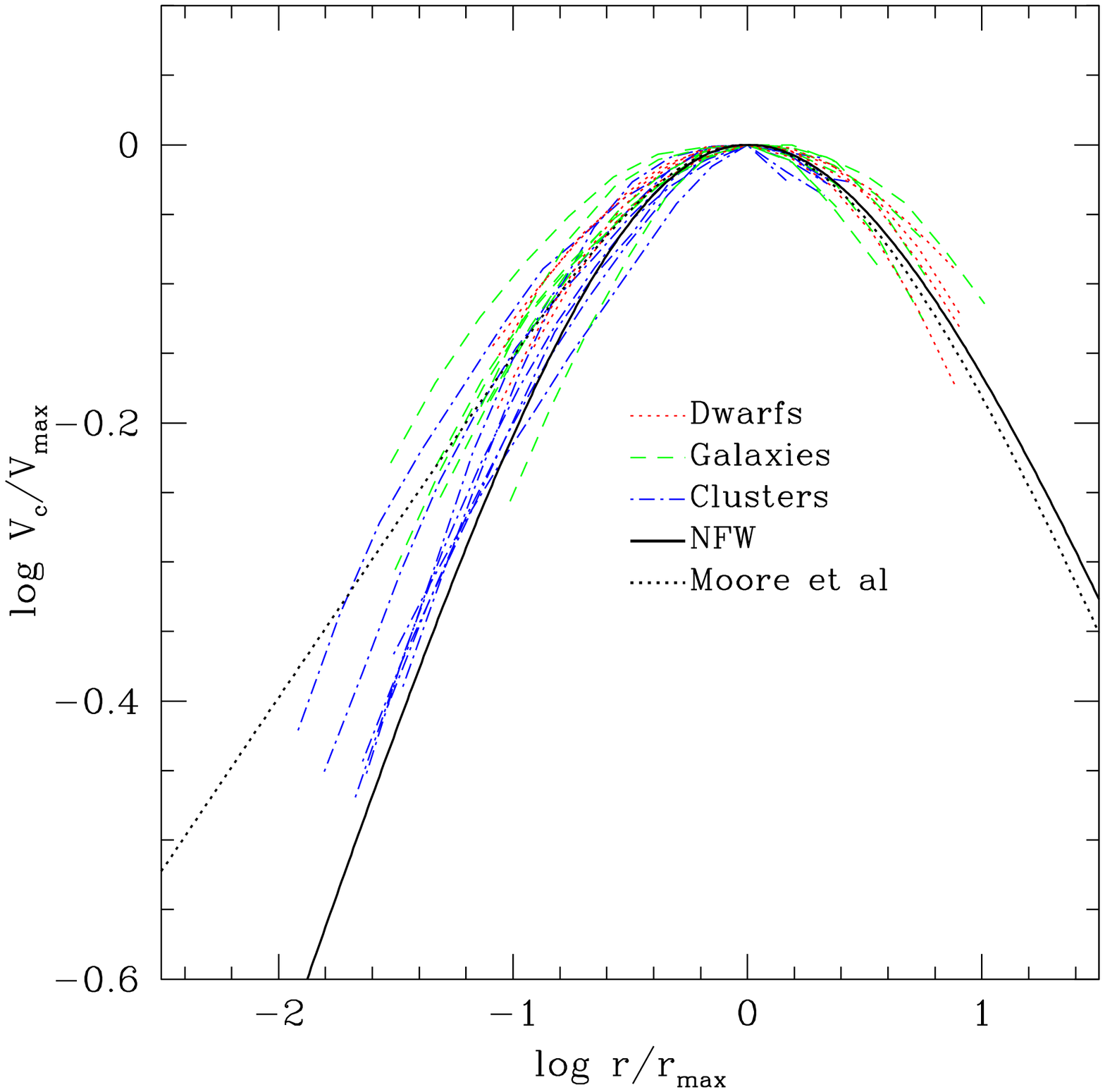}
\end{center}
\caption{\small {\it (a)} Density profiles of 19 $\Lambda$CDM halos, scaled
to the radius where the logarithmic slope takes the ``isothermal'' value of
$\beta=-d\log\rho/d\log r=2$.  Thick solid curves show the NFW profile; dotted
curves correspond to the M99 profile. {\it (b)} Circular velocity profiles,
scaled to the peak of the $V_c$ curve. Note that the NFW and M99 profile appear
to ``bracket'' the most extreme mass profile shapes of $\Lambda$CDM halos.}
\end{figure}

Figure 1 shows the density and circular velocity profiles of nineteen
$\Lambda$CDM halos simulated in our series. The ``universality'' in
the density profile shape is displayed by scaling each profile to
$r_{-2}$, the radius where the logarithmic slope is
$\beta=-d\log\rho/d\log r=2$. The left panel of Figure 1 shows that
there is little difference in the scaled density profiles of CDM halos
differing by up to five decades in mass. The same applies to the $V_c$
profiles, which in the right panel of Figure 1 have been scaled to the
peak in the circular velocity curve. The fitting formulae proposed by
NFW or M99 are fixed curves in these scaled units, and are seen to
more or less ``bracket'' the most extreme shapes of the mass profiles
of the simulated halos.

Figure 2a shows the radial dependence of $\beta$ in the simulated halos, and
confirms that neither the NFW profile nor the M99 profile capture fully the
structural diversity of $\Lambda$CDM halos. The profiles are clearly shallower
near the center than the asymptotic value of $\beta_0=1.5$ proposed by M99. At
the same time, $\beta(r)$ deviates from the NFW prediction inside $r_{-2}$:
density profiles become shallower less rapidly than expected from the NFW
formula.  Thus, NFW fits tend to underestimate systematically the density just
inside $r_{-2}$ (see Figure 1a). Fitting formulae with steeper cusps (such as
the M99 profile) might indeed do better in that region; however, so will any
other modification of the NFW formula where the logarithmic slope inside
$r_{-2}$ depends more weakly on radius than NFW's , {\it regardless} of
$\beta_0$.

\begin{figure}
\begin{center}
\plottwo{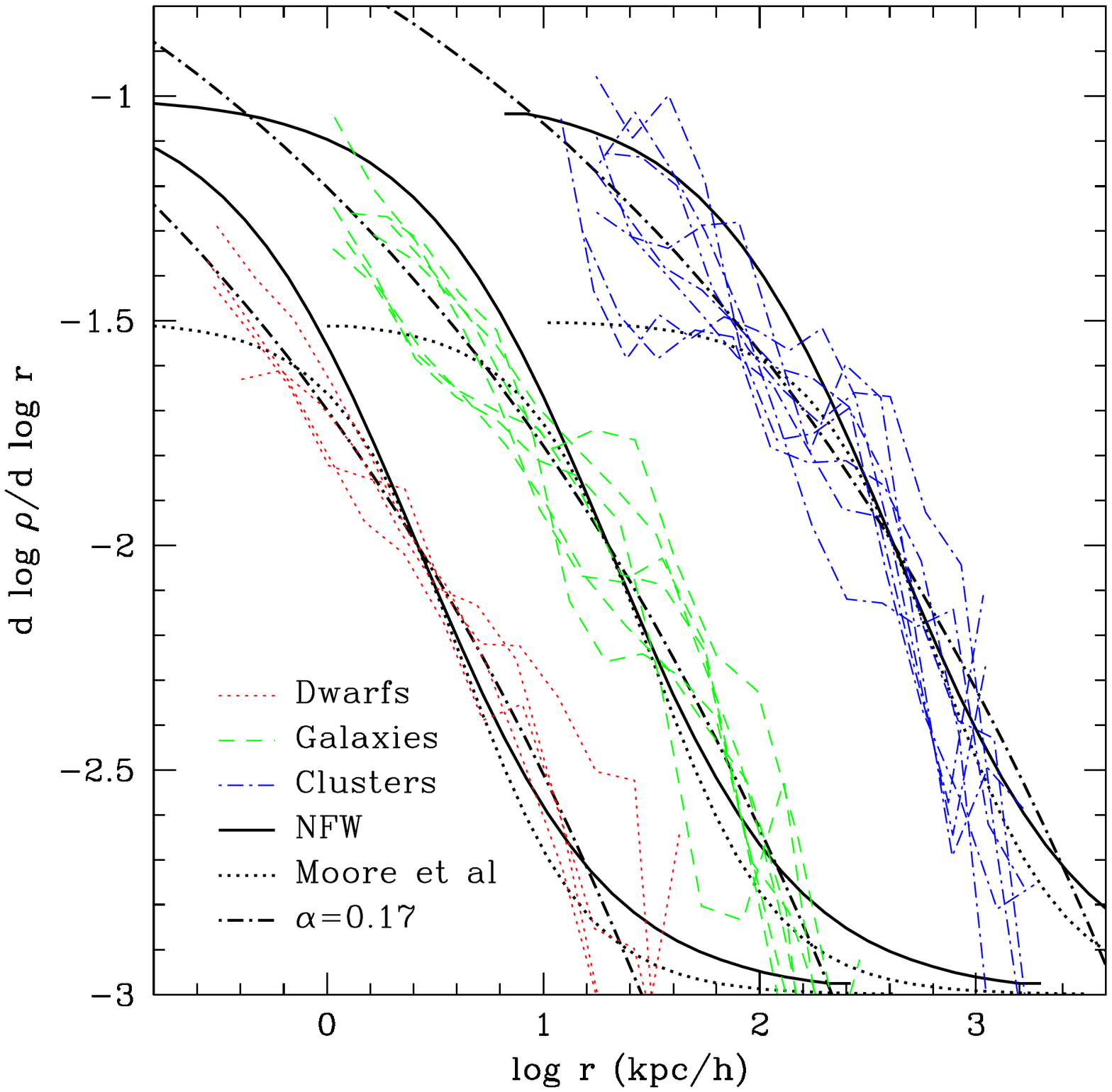}{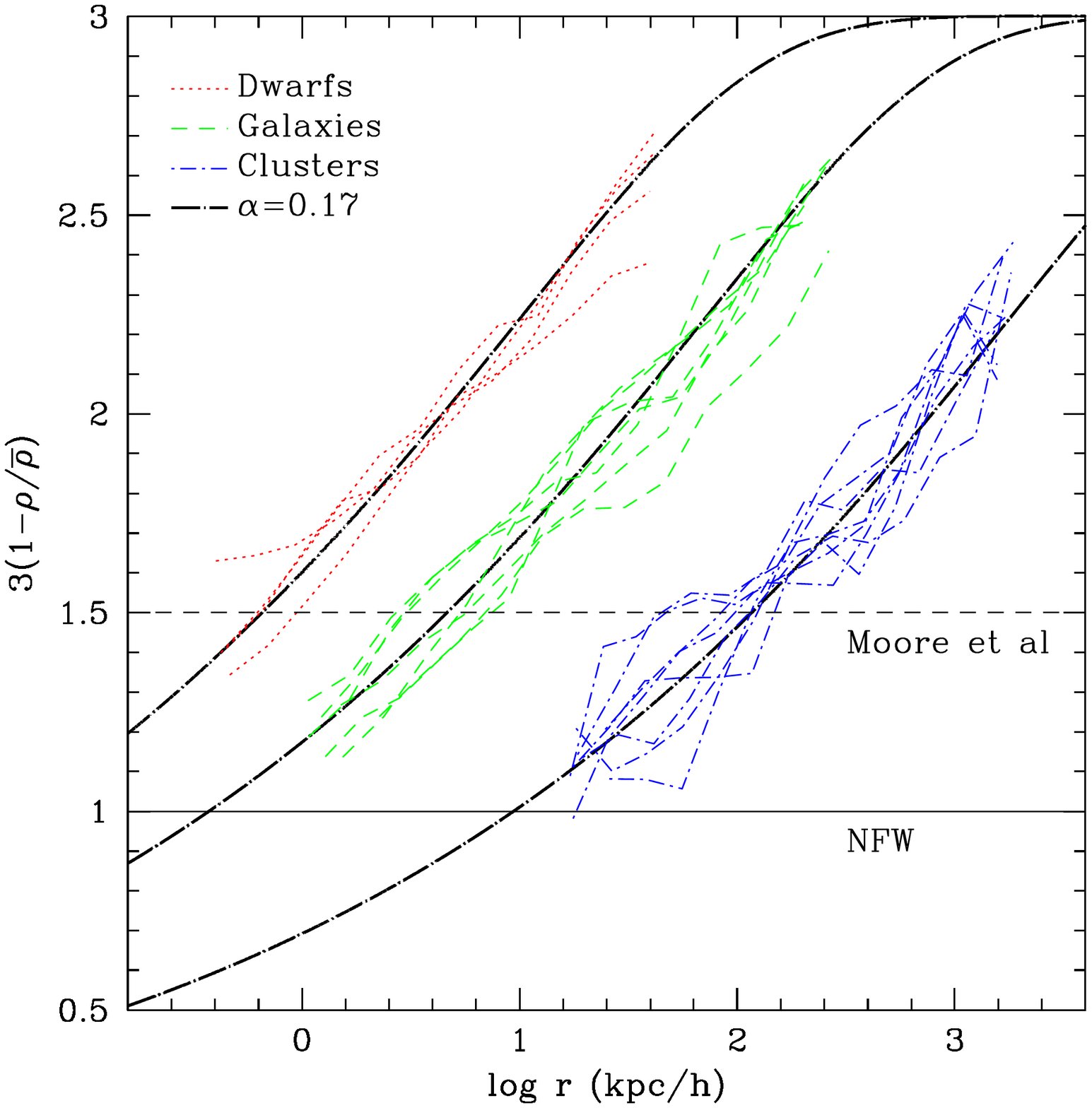}
\end{center}
\caption{\small {\it (a)} Logarithmic slope of the density profile as a
function of radius for (from left to right) ``dwarf'',
``galaxy-sized'', and ``cluster-sized'' $\Lambda$CDM halos. The radial
dependence of the slope is significantly different from both the NFW
and M99 profiles, and is better approximated by the $\rho_{\alpha}$
profile described in eqs.(2) and (3). {\it (b)} Maximum inner
asymptotic slope as a function of radius. Cusps as steep as $\rho
\propto r^{-1.5}$ are clearly ruled out by our simulations;
$\beta_0=1$ (as in the NFW profile) are still consistent with the
data, although there is little evidence for convergence to a power law
density profile at the center. }
\end{figure}

Conclusive proof that the cusps cannot diverge as steeply as $r^{-1.5}$ is
provided by the total mass inside the innermost resolved radius, $r_{\rm conv}$
(the minimum radius plotted for all profiles shown in this paper). This is
because the mean density interior to {\it any} radius, $\bar
\rho(r)$, together with the local density at that radius, $\rho(r)$, provide a
robust limit to the asymptotic central slope, $\beta_0<\beta_{\rm
max}(r)=3(1-\rho(r)/{\bar \rho(r)})$ (under the plausible assumption that
$\beta$ is monotonic with radius). The radial dependence of $\beta_{\rm max}$ is
shown in figure 2b, and shows that, except for possibly one dwarf system, no
simulated halo can have a cusp as steeply divergent as $\beta_0=1.5$. 

Actually, there is no indication in the profiles for a well defined value for
$\beta_0$, as profiles keep getting shallower down to the smallest resolved
radius.  A simple power law approximates the radial dependence of $\beta(r)$
better than the NFW or M99 profiles;
\smallskip
\begin{equation}
\beta_{\alpha}(r)=-{d\ln \rho/d\ln r}=2\left({r/r_{-2}}\right)^{\alpha},
\label{eq:newbeta}
\end{equation}
\smallskip
which corresponds to a density profile of the form,
\smallskip
\begin{equation}
\ln (\rho_{\alpha}/\rho_{-2})={(-2/\alpha)} [\left({r/r_{-2}}\right)^{\alpha} -1].
\label{eq:newprof}
\end{equation}
\smallskip
The thick dot-dashed curves in Figure 2 show that eq.2 (with $\alpha\sim 0.17$)
does indeed reproduce fairly well the radial dependence of $\beta(r)$ and
$\beta_{\rm max}(r)$ in simulated halos. Furthermore, adjusting the parameter
$\alpha$ allows the profile to be tailored to each individual halo, resulting in
much improved fits over those obtained with the NFW or M99 formulae.  The
best-fit values of $\alpha$ (in the range $0.1$ - $0.2$) show no obvious
dependence on halo mass: the average $\alpha$ is $0.172$ and the dispersion
about the mean is $0.032$ for the nineteen halos in our series (see Navarro et
al 2003 for further details).

\section{CDM halos and LSB rotation curves}

A number of authors have reported disagreements between the shape of rotation
curves of low surface brightness (LSB) galaxies and the circular velocity
profiles implied by fitting formulae such as that proposed by NFW (Moore 1994,
Flores \& Primack 1994, McGaugh \& de Blok 1998, de Blok et al 2001). Many of
these galaxies are better fit by circular velocity curves arising from density
profiles with a well defined constant density ``core'' rather than the cuspy
ones inferred from simulations, a result that has prompted calls for a radical
revision of the CDM paradigm on small scales (see, e.g., Spergel \& Steinhardt
2000). However, before accepting the need for radical modifications to CDM it is
important to note a couple of caveats that apply to the LSB rotation curve
problem.

\begin{itemize}

\item
Strictly speaking, the observational disagreement is with the fitting formulae,
rather than with the actual structure of simulated CDM halos.  As noted in the
previous section, there are small but systematic differences between them, so it
is important to confirm that the disagreement persists when LSB rotation curves
are contrasted directly with simulations.

\item
It must be emphasized that the rotation curve problem arises when comparing
rotation speeds of LSB disks to spherically-averaged circular velocities of dark
matter halos. Given that CDM halos are expected to be significantly
non-spherical, some differences between the two are to be expected. It is
therefore important to use the full 3D structure of CDM halos to make
predictions regarding the rotation curves of gaseous disks that may be compared
directly to observation.

\end{itemize}

Figure 3a illustrates the LSB rotation curve problem highlighted above. This
figure shows the rotation curves of four LSB galaxies (points with error bars)
selected from the sample of McGaugh et al (2001). The data points have been
fitted using a simple formula, $V_{\rm rot}(r)=V_0
(1+(r/r_0)^{-\gamma})^{-1/\gamma}$ (Courteau 1997). Here $V_0$ and $r_0$ are
dimensional scaling parameters, whereas $\gamma$ is a dimensionless parameter
that characterizes the {\it shape} of the rotation curve. This three-parameter
formula provides excellent fits to all galaxies, as witnessed by the quality of
the (solid line) fits shown in figure 3a.

\begin{figure}
\begin{center}
\plottwo{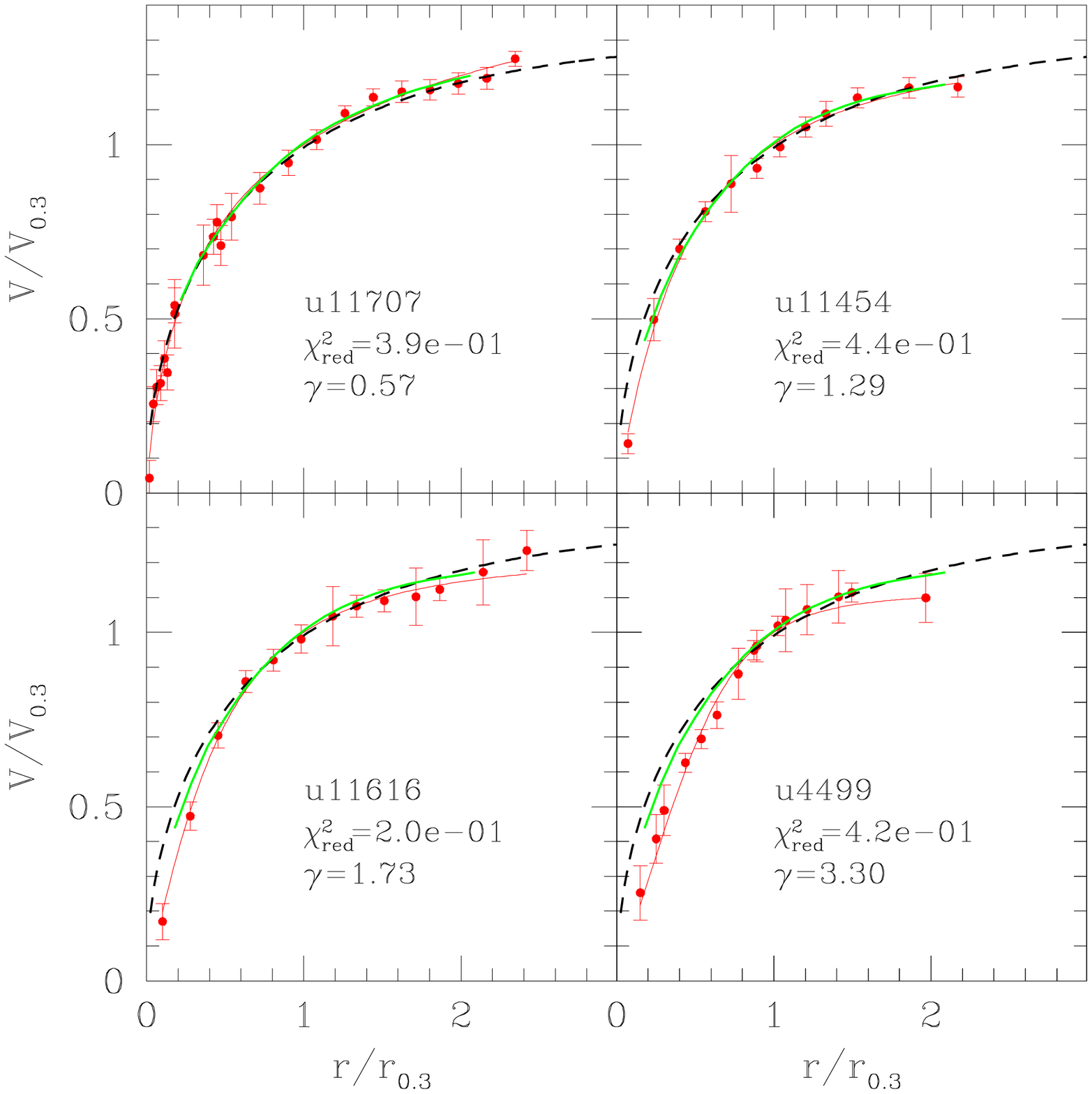}{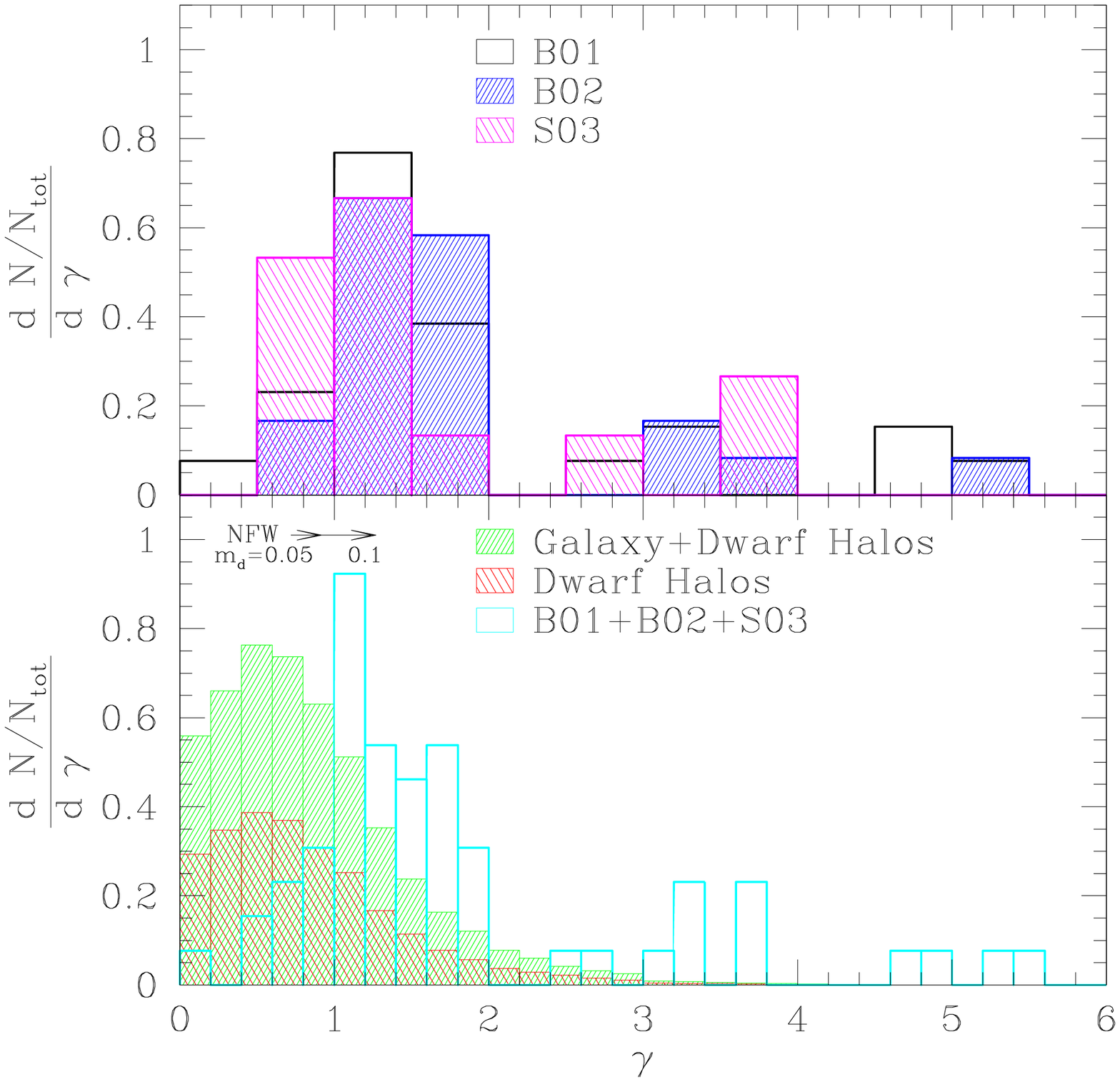}
\end{center}
\caption{\small {\it (a)} Rotation curves of four LSB galaxies chosen to
illustrate their various shapes, as measured by $\gamma$. The larger
$\gamma$ the sharper the turn of the rotation curve from rising to
flat, and the poorer the agreement with NFW profiles (dashed
lines). The $V_c$ profile of simulated halos (solid lines) match
reasonably well systems with $\gamma \la 2$ but cannot account for
those with $\gamma>2$.{\it (b)} Top panel shows the distribution of
the ``shape'' parameter $\gamma$ corresponding to all rotation curves
in the samples of de Blok et al (2001, 2002) and of Swaters et al
(2003). Bottom panel compares the distribution of ``observed''
$\gamma$ with that derived from fits to halo circular velocities. This
shows that it is possible to account for rotation curves with
$\gamma\la 2$ but that curves with $\gamma \ga 2.5$ are inconsistent
with halo $V_c$ profiles.  }
\end{figure}

To emphasize visually the shape discrepancy, the rotation curves in figure 3a
have been scaled to $r_{0.3}$ and $V_{0.3}$: the radius and velocity,
respectively, where the slope of the curve is dln$V_{\rm rot}$/dln$r=0.3$. The
four galaxies have different values of $\gamma$, and have been chosen to
illustrate to whole range of rotation curve shapes in the McGaugh et al and
Swaters et al datasets. Roughly two-thirds of their LSBs have $0.5<\gamma<2$;
the rest have $2<\gamma<5$ (see top panel of Fig.3b).

The dashed lines in figure 3 show the $V_c$ profile of an NFW halo,
which is fixed in these scaled units. Clearly, as $\gamma$ increases,
the difference in shape between the NFW profile and the rotation curve
becomes more acute. Galaxies with $\gamma \la 2$ are roughly
consistent with NFW, whereas $\gamma>2$ rotation curves are clearly
inconsistent. Can the deviations from NFW discussed in figures 1 and 2
be responsible for this discrepancy?

To address this question we have fitted all simulated halos with the
same ($V_0,r_0,\gamma$) formula used for the observational data. Each
halo is fitted at various times during its late evolution so as not to
impose any restrictions on the dynamical state of the halo; temporary
departures from equilibrium associated with recent accretion events
are thus taken into account `naturally'' in this procedure.  
%
The $\gamma$ distribution of halos peaks at about $\gamma\sim 0.6$ and is
restricted to values that rarely exceed $\sim 2.5$ (see bottom panel of Fig.3b).
Halos whose fitted $\gamma$ best match each of the galaxies in Figure 3a are
shown with faint solid lines in that figure.


Full details of this exercise are presented in Hayashi et al (2003), but the
main conclusion is that it is possible to find halos whose $V_c$ curves match
all galaxies with $\gamma \la 2$, with scaling parameters $r_0$ and $V_0$ in
reasonable agreement with observed values.  {\it CDM halos are thus actually
consistent with the majority of LSB rotation curve data}. Indeed, the rotation
curve problem may be thought of as restricted to the few galaxies with
$\gamma>2$; these rotation curves rise and turn too sharply to be consistent
with the $V_c$ profiles of simulated halos (see the bottom right panel of figure
3a).

Do these galaxies rule out the presence of a cusp in the dark matter
density profile? As noted above, before concluding so we must take
into account possible systematic differences between rotation speed
and circular velocity in gaseous disks embedded within realistic,
triaxial halos. This is a complex issue that involves a number of
parameters, such as the degree of triaxiality, the role of the disk's
self-gravity, size, and orientation, etc.

For the sake of simplicity, we have decided to begin addressing this issue by
evolving a massless, isothermal gaseous disk at the center of mildly triaxial
(1:0.9:0.8) halos (Hayashi et al, in preparation). The triaxiality in the
potential induces a radially-dependent inclination which may fluctuate, in the
case we consider here, by $5$ to $10$ degrees from one radius to another. The
effect on rotation curves of departures from circular symmetry and from coplanar
orbits derived from long-slit spectra is complex, but the {\it shape} of the
rotation curve, in particular, is affected. On some projections, rotation curves
appear to rise and turn rapidly, and they would be (erroneously) taken to imply
the existence of a constant-density core in simple models that impose spherical
symmetry. Fits to the rotation curves using the ($V_0,r_0,\gamma$) formula often
have $\gamma>2$, consistent with galaxies where NFW profiles provide a
particularly poor fit to the rotation curve data. This suggests that deviations
from spherical symmetry in the mass structure of CDM halos might reconcile
rotation curve shapes that seem to favor the presence of constant density cores
with cusps in the dark matter density profiles.

\section{Concluding Remarks}

Although it appears possible to reconcile dark matter cusps with LSBs
by appealing to asphericity in CDM halos, it would be premature to
argue that the problem has been fully solved. After all, it may come
as no surprise that one is able to reproduce LSB rotation curves,
given the number of extra ``free'' parameters afforded by relaxing the
assumption of spherical symmetry. It is therefore important to build a
more compelling case for this interpretation of rotation curve data,
so as to render it falsifiable. We are in the process of identifying
corroborating trends that may be used either to support or to rule out
this interpretation, as the case may be. In particular, we would like
to characterize better the $\gamma>2$ systems: are such rotation
curves in general asymmetric?  Is rotation around the minor axis
expected? Can one verify the triaxial-halo interpretation in full
velocity maps? In this sense, identifying a ``make or break''
prediction will be as important as the success of aspherical halos in
reproducing the rich variety of shapes of LSB rotation curves. Only
once this is accomplished shall we be able to conclude that LSB
rotation curves do not preclude the presence of dark matter density
cusps, freeing the CDM paradigm of one of its most vexing challenges
on small scales.

\acknowledgements I am grateful to my collaborators, Eric Hayashi;
Adrian Jenkins; Carlos Frenk; Simon White; Volker Springel; Chris
Power; Thomas Quinn; and Joachim Stadel, for allowing me to report
some of our results in advance of publication. This work has been
supported by the Alexander von Humboldt Foundation, NSERC and CFI, and
by fellowships from CIAR and from the J.S.Guggenheim Memorial
Foundation.


\begin{references}

\reference
Courteau, S.\ 1997, AJ, 114, 2402

\reference
Crone, M., Evrard, A.E., \& Richstone, D.O. 1994, ApJ, 434, 402.

\reference
de Blok, W.~J.~G. et al 2001, ApJL, 552, L23

\reference
de Blok, W.~J.~G, Bosma, A.  2002, AAP, 122, 385, 816.

\reference
Dubinski, J. \& Carlberg, R. 1991, ApJ, 378, 496.

\reference
Flores, R., \& Primack, J.R. 1994, ApJL, 427, L1.

\reference
Frenk, C.S., et al 1988, ApJ, 327, 507.

\reference
Fukushige, T.~\& Makino, J.\ 2001, ApJ, 557, 533

\reference
Fukushige, T.~, Kawai, A.~\& Makino, J.\ 2003 (astro-ph/0306203)

\reference
Jing, Y.\ P.\ \& Suto, Y.\ 2000, ApJL, 529, L69

\reference
Hayashi, E. et al 2003, MNRAS, submitted (astro-ph/0310576)

\reference
Klypin, A.A. et al 2001, ApJ, 554, 903.

\reference
McGaugh, S.S., De Block, W.J.G. 1998, ApJ, 499, 41.

\reference
McGaugh, S.~S., Rubin, V.~C., de Blok, W.J.G., 2001, AJ, 122, 2381.
552, L23

\reference
Moore, B. 1994, Nature, 370, 629.

\reference
Moore B., Governato F., Quinn T., Stadel J., Lake G., 1998, ApJ, 499, L5

\reference
Moore, B. et al, 1999, ApJL, 524, L19

\reference
Navarro, J. F., Frenk, C. S. \& White, S. D. M. 1996, ApJ, 462, 563

\reference
Navarro, J. F., Frenk, C. S. \& White, S. D. M. 1997, ApJ, 490, 493

\reference
Navarro, J.F. et al 2003, MNRAS, submitted (astro-ph/0311231)

\reference
Power, C., et al. 2003, MNRAS, 338, 14.

\reference
Spergel, D., Steinhardt, P., 2000, PRL, 84, 3760.

\reference Swaters, R.A. et al 2003, ApJ 583, 732.

\reference
Taylor, J.E., \ \& Navarro, J.F. 2001, ApJ, 563, 483.

\reference
van den Bosch, F. et al 2000, AJ 119, 1579.

\end{references}
\end{document}